\documentstyle[12pt,epsf,rotate]{article}
\begin{document}

\hfill{\sf Gamma-Ray Burst in the Afterglow Era: 2nd Workshop}

\hfill{\sf Rome, Italy, October 17-20, 2000}

\vspace*{0.5cm}

\centerline{\huge\bf Wolf-Rayet stars and GRB connection}


\begin{center}

{\Large
Anatol Cherepashchuk,
Konstantin Postnov \\
}
{\it Sternberg Astronomical Institute, Moscow State University, 119899
Moscow, Russia} 

\end{center}


\begin{abstract}
{\large\sf\noindent Arguments are given favoring possible connection of GRBs
with core collapse of massive Wolf-Rayet stars. We analyze the observed
properties of cosmic gamma-ray bursts, Wolf-Rayet (WR) stars and their
CO-cores in the end of evolution. WR stars are deprived of their extended
hydrogen envelopes, which makes it easier for the collapse energy to
transform into observed gamma-ray emission. Presently, of $\sim 90$
well-localized gamma-ray bursts, 21 ones are optically identified and for 16
of them redshifts are measured ($z=0.4\div 4.5$). The observed energy of
gamma-ray bursts spans over a wide range from
$3\times 10^{51}$ to $2\times 10^{54}$ ergs. 
There is some evidence that this distribution 
$N(\Delta E)$ is bimodal if take into account GRB980425 associated with a
peculiar type Ic supernova SN1998bw in a nearby galaxy ESO 184-G82 for which
$\Delta E_\gamma\approx 10^{48}$ ergs.
These characteristics of gamma-ray bursts are similar to the distribution of
the final masses of CO-cores of WR stars which is also wide and homogeneous:
$M_{CO}= (1-2)M_\odot\div (20-44)M_\odot$. 
A possible bimodality of the gamma-ray burst energy distribution
($E_1=10^{48}$ erg; $\Delta E_2=3\times 10^{51}\div 2\times 10^{54}$ erg) is
in accord with the bimodal mass distribution of relativistic objects
($M_{NS}=(1.35\pm 0.15) M_\odot$; $M_{BH}=(4\div 15) M_\odot$). That the
supernova SN1998bw is of the "peculiar Ic" type, atypical for WR collapses
(type Ib/c), can be related to the rotation of the collapsing CO-core which
can make the collapse longer and lead to the formation of a neutron star,
the decrease of the gamma-ray burst energy, and the increase of the fraction
of kinetic energy transported to the envelope. The expected collapse rate of
CO-cores of most compact WR stars of type WO in the Galaxy is $\sim 10^{-5}$
per year, which is only by one and a half order of magnitude higher than the
average gamma-ray burst rate in one galaxy ($\sim 10^{-6}-10^{-7}$ per
year). Two particular models that use WR stars as gamma-ray burst
progenitors are considered: the hypernova model by Paczy\'nski (1998) and
the model of unstable CO-core collapse suggested by Gershtein (2000). In
both models the allowance of a gamma-ray beaming or random outcome of the
CO-core collapse due to some instabilities permits one to bring the rate of
CO-core collapses in accordance with that of gamma-ray bursts. We argue that
WR stars (most probably, of type WO) can be considered as progenitors of
cosmic gamma-ray bursts. Two types of gamma-ray bursts are predicted in
correspondence with the bimodal mass distribution of the relativistic
objects. Three types of optical afterglows should appear depending on which
CO-core is collapsing: of a single WR star, of a WR star in a WR+O or a
hypothetic WR+(A-M) binary system. In addition, we briefly consider a model
of gamma-ray bursts as a transient phenomenon occurring at early stages of
galactic evolution ($z>1$), when very massive ($M>100 M_\odot$)
low-metallicity stars could form. Such massive stars should also loose their
hydrogen envelopes and become massive WR stars whose collapses could be
accompanied by gamma-ray bursts. WR-galaxies can be most probable candidates
for gamma-ray burst host galaxies.
}

\end{abstract}

\vspace*{0.2cm}

\baselineskip+20pt
\section*{Introduction}

The principal requirements to the central engine of GRBs include
(1) The ability to release the electromagnetic energy $\sim
10^{52}$ ergs during 10-100 s (the typical duration of "long" GRBs, only for
which X-ray and optical observations are possible; a separate group of
short single-peak GRBs is much less studied, apart from the fact 
that they are isotropically
distributed over the sky, and well may be another phenomenon) and (2)
The event rate is on average about one burst per typical galaxy (assuming
isotropy of the emission and homogeneity of galaxies) per $\sim 10^7$
years. The beaming of
gamma-ray emission decreases the energy emitted and increases by the same
amount the event rate.

These requirements are met (with different degree of accuracy) by
two broad classes of astrophysical sources. The first class includes
coalescencs of binary NS and/or BH
\cite{Blinnikov_ea84}).  The fireball is generated by 
neutrino-antineutrino annihilation copiously produced 
during the coalescence.  The second class comprises models
related to final stages of evolution of massive stars, among which:

\begin{itemize}

\item
Collapses of massive stars \cite{Woosley93,Paczynski97,Macfadyen&Woosley99}.  
An accretion disk is formed around a 
massive rotating black hole during late stages of 
the core collapse of a massive star; in this model, 
a narrow jet is produced inside the star and punches
through the stellar envelope reaching very high Lorentz-factors.

\item
Electromagnetic model by V.Usov 
\cite{Usov92}, in which the energy comes from the rapid rotation of 
a young neutron star (millisecond pulsar) with a very strong
magnetic field. Other models in which magnetic field is crucial 
see in \cite{Vietri&Stella98, Spruit99}.

\item
GRB during core collapse of a non-rotating Wolf-Rayet star
by Gershtein \cite{Gershtein00}, in which the internal shocks are
created due to the collapse non-stationarity and energy 
is brought away by electron-positron plasma. 

\end{itemize}

At the late stages of evolution, very massive stars
lose their hydrogen-rich envelopes (via stellar wind or 
in a binary system) and are observed as
helium-rich Wolf-Rayet (WR) stars. They are considered as progenitors
of Ib type supernovae (if strong helium lines are present 
in the maximum brightness) or Ic supernova (if helium lines
are weak or absent).  If GRBs are directly related to collapses
to massive stars, there should be some links between properties 
of WR stars before collapse and GRBs. We suggest that observed broad 
(and, possibly, bimodal) distribution 
of GRB energetics is related (1) to  the broad distribution of 
final masses of CO cores of {\it observed} WR stars before collapse and 
(2) to the {\it observed} bimodal distribution of masses
of relativistic remnants of stellar evolution
(neutron stars and black holes).

\section*{GRB association with star formation}

There is a growing evidence that cosmic GRBs are associated 
with star-forming regions:

\begin{itemize}

\item
Optical observations of the identified GRB host galaxies 
evidence for an enhanced star formation rate
inside the GRB sites, sometimes by an order of magnitude higher
than in our Galaxy \cite{Vreeswijk_ea00}. 

\item
Large column densities  $10^{22}-10^{23}$~cm$^{-2}$
toward GRBs derived from 
known X-ray and optical afterglows of GRBs
\cite{Galama&Wijers00}, typical for giant
molecular clouds. 

\item
Observed offset distribution of GRBs from their host galaxies
statistically suggests association with the  location of massive stars
\cite{Bloom_ea00.astro-ph/0010176}.

\item
Detections of line features in several GRB X-ray afterglows
(\cite{Antonelli_ea00.astro-ph/0010221} and references therein)
imply a dense surrounding.

\end{itemize}

This favors 
GRBs association with evolution of massive stars and makes
other GRB progenitors (such as double neutron star 
coalescences) less likely.

\section*{Energetics of GRBs with known redshifts}
{\large
\begin{table}
\caption{GRBs with known energetics}
\label{GRB}
\begin{tabular}{lccccccc}
\hline
\hline            
GRB            & z     & $d_l^\dag$,  $10^{28}$& $F_\gamma$, $10^{-5}$ & Ref& $\Delta E$, $10^{53}$ & $F_p^\ddag$  & $L_p$, $10^{58}$ \\
               &       & cm& erg/cm$^2$           &     & erg& ph/s/cm$^2$ & ph/s \\
               &       &             &(10-2000 keV)         &     &             & (50-300
keV) \\
\hline
000926         & 2.066:& 5.81     &2.2  &\protect{\cite{JPUFynbo_ea00GCN807}}& 3.04
\\
000418$^{a)}$  & 1.118 & 2.73     &1.3  & \protect{\cite{JBloom_eaGCN661}}       &$\sim 0.6$ &  -    \\
000301C$^{b)}$& 2.03  & 5.69     &$>0.05$ & \protect{\cite{SMCastro_eaGCN605}}     &$\sim 0.07$& $\sim 5$ & 6.7    \\
991208$^{c)}$& 0.706 & 1.55     &10   & \protect{\cite{SGDjorgovski_eaGCN481}} &$\sim 1.8 $&  -     \\
990712         & 0.430 & 0.85     &-      \\
990510	       & 1.619 & 4.30     &2.26 & \protect{\cite{Kumar&Piran00}}         &2.0        &8.16& 7.3    \\
990123	       & 1.6   & 4.25     &26.8 & \protect{\cite{Kumar&Piran00}}         &23         &16.4& 14    \\    
980703         & 0.967 & 2.28     &2.26 & \protect{\cite{Kumar&Piran00}}         &0.75       &2.6& 0.86  \\
980613$^{d)}$  & 1.096 & 2.66     &0.17 & \protect{\cite{Briggs_ea99}}
&0.072         &0.63& 0.27  \\
971214         & 3.412 & 10.6     &0.944& \protect{\cite{Kumar&Piran00}}         &3.0        &2.3 & 7.4  \\
970828	       & 0.958 & 2.25     &9.6  & \protect{\cite{Kumar&Piran00}}         &3.1        & - \\
970508	       & 0.835 & 1.99     &0.317& \protect{\cite{Kumar&Piran00}}         &0.08       &1.2& 0.29  \\
970228$^{d)}$  & 0.695 & 1.52     &$\sim 0.2$& \protect{\cite{Briggs_ea99}}
&0.034          &3.5& 0.60  \\
\hline
980425         & 0.0085 & 0.013   & 0.32 & \protect{\cite{Galama_ea98}} &
$7\cdot 10^{-6}$ & 0.96 & $2.1\cdot 10^{-5}$\\
\hline
\end{tabular}
\medskip

Notes:

$\dag$ Flat Universe, $\Omega_m=0.3, \Omega_\Lambda=0.7, H_0=60$~km/s/Mpc 

$\ddag$ Peak fluxes from \protect{\cite{Lamb00}}  

a) Energy fluence in 25-100 keV;

b) Peak flux $F_p=3.7$ Crab, no fluence published, 
single-peak profile, duration 10 s; other indirect estimations of
total energy in gamma-rays see in \protect{\cite{Sagar_ea}} 

c) Energy fluence for $E>25$ keV

d) Energy fluence for $E>20$ keV
\end{table}
}

Table \ref{GRB} lists GRBs with known energetics. 
Photometric distances
are calculated using a flat Universe with $\Omega_m=0.3,
\Omega_\Lambda=0.7, H_0=60$~km/s/Mpc. The Table also contains  
BATSE fluences (50-300 keV) and peak luminosities $L_p$ phot/s
(from \cite{Lamb00}). 

\begin{figure}
\centerline{
\hss
\epsfysize=7cm
\epsfbox[100 110 500 650]{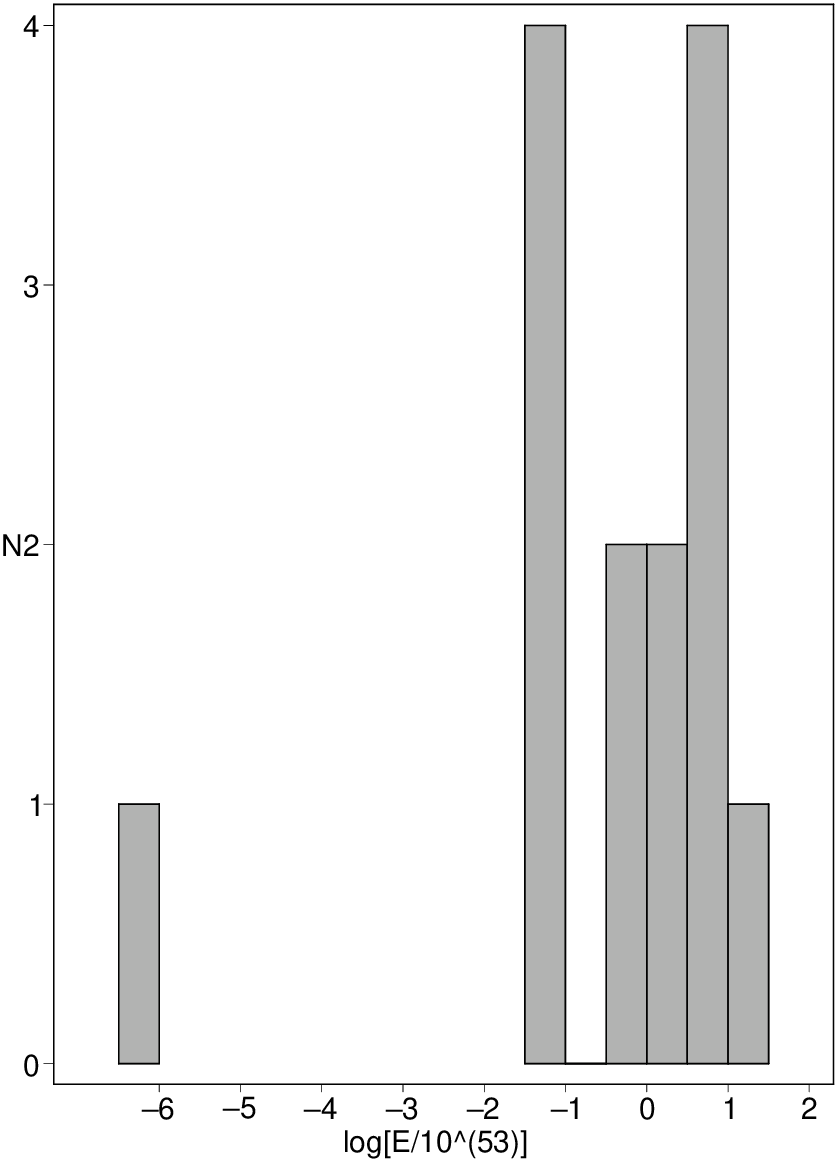}
\hss
\epsfysize=7cm
\epsfbox[100 110 500 650]{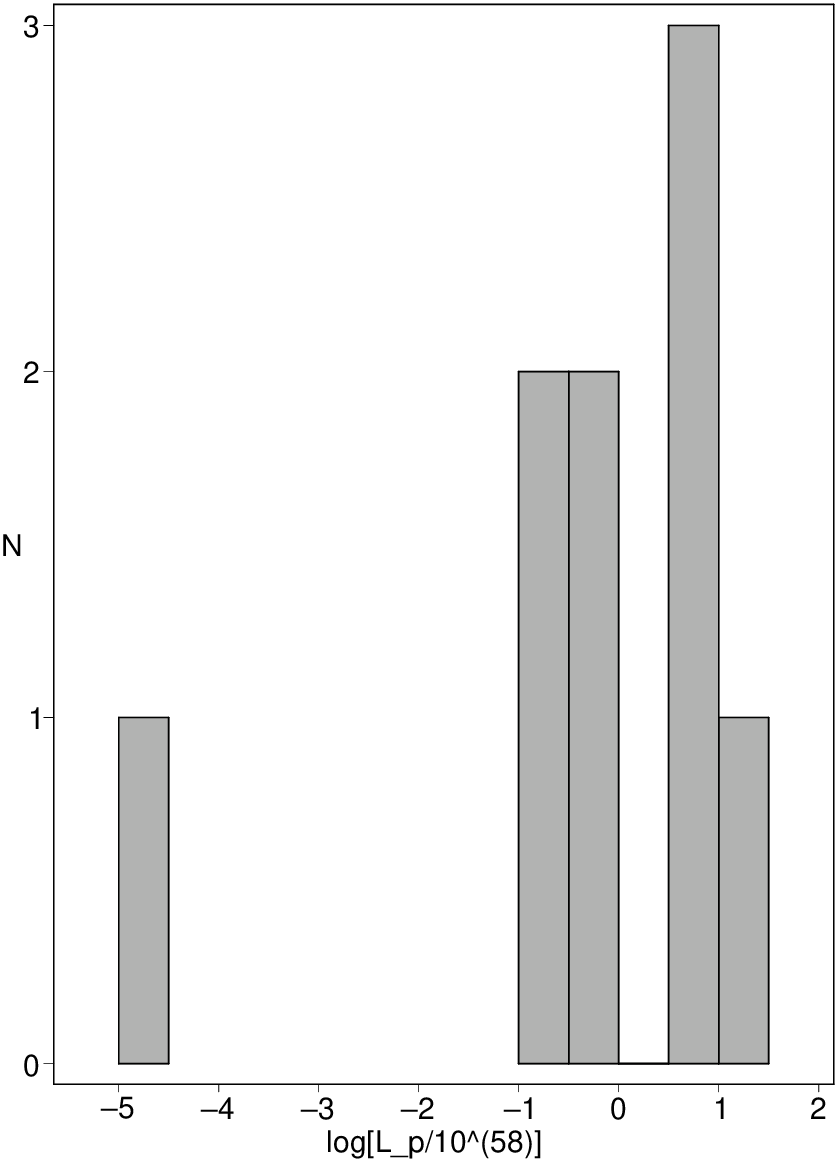}
\hss
}
\caption{Total GRB energy release $\Delta E$ 
(in $10^{53}$ erg/s) of GRBs with known redshifts 
from Table \protect\ref{GRB} }
\label{energy}
\caption{Peak luminosities $L_p$ (in 
$10^{58}$ phot/s) 
of GRBs from Table \protect\ref{GRB}}
\label{flux}
\end{figure}

As seen from Table
\ref{GRB} and Fig. \ref{energy}, \ref{flux},
the observed GRB energetics spans from  
$\approx 7\times 10^{51}$ ergs to  $\approx 2\times  10^{54}$ ergs. 
This observational fact is usually explained by 
a broad luminosity function of GRBs
(see \cite{Loredo&Wasserman98}, \cite{Schmidt00}),
although one can construct a self-consistent model with
a universal energy release of 
$E_0\sim 5\times 10^{51}$ ergs and a complex beam shape  
\cite{LPP00}. Adding GRB980425, which is possibly associated 
with peculiar type Ic supernova SN 1998bw in a nearby ESO 184-G82 
($z=0.0085$) \cite{Galama_ea98}, 
either evidences
for a {\it bimodality} of GRB energy distribution 
or for 
an extremely broad luminosity function (more than 5 orders of magnitude!).

\section*{GRB and WR stars} 

The arguments 
favoring the link with WR stars 
are \cite{Postnov&Cherepashchuk00}:
\vskip\baselineskip

\begin{figure}[t]
\centerline{
\epsfysize=\hsize
\fbox{\rotate{\epsfbox{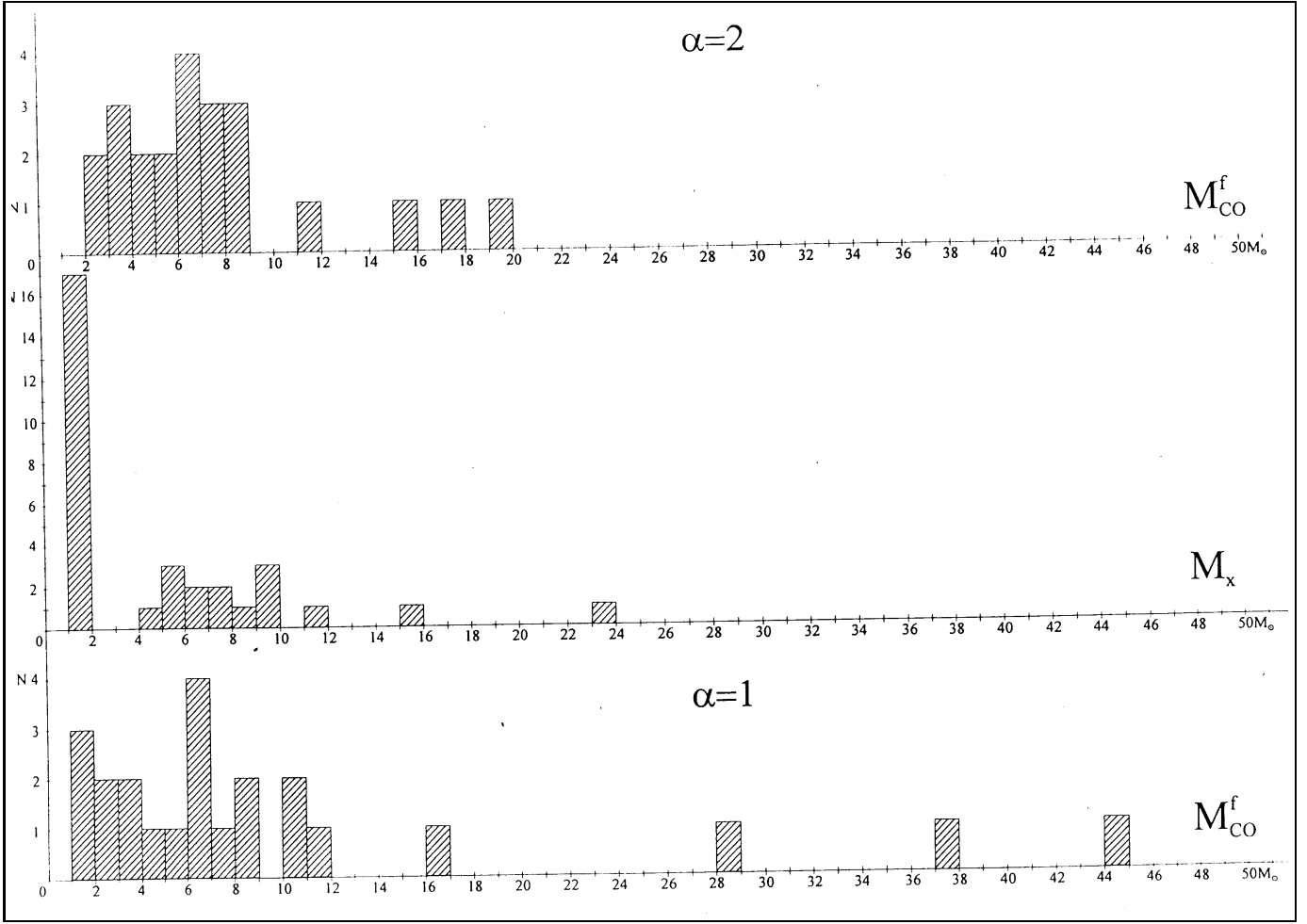}}}
}
\caption{Masses of final CO cores $M_{CO}$ of
WR stars as calclated using observed 
stellar wind mass loss $\dot M\propto M^\alpha$. 
Masses of known NS and BH candidates in binary systems
are also shown.}
\label{masses}
\end{figure}

(1) {\bf Energetics}
The observed GRB energy release 
$\Delta E\simeq 10^{51}-10^{54}$ erg is roughly comparable
with the wide range of CO cores of WR stars before collapse, 
$M_{CO}^f\simeq 2-40 M_\odot$, Fig \ref{masses}. 
The total energy released
in collapse is $E_G\sim GM_c^2/R_c$, where $M_c$ in the compact remnant
mass, $R_c$ its radius. During black hole formation in such collapses
without mass ejection the available energy can reach
$10^{53}-10^{56}$ ergs for the observed mass range of
compact remnants, i.e. conversion of 1\% of the available
energy into kinetic energy of shocks with subsequent radiation would be
sufficient to explain the broad luminosity function. Note that during
collapse into black hole without mass ejection
$R_c\propto M_c\propto M_{CO}$ and available energy range is proportional 
to the CO core mass range, which is likely to be the case.

\begin{figure}[t]
\centerline{
\epsfysize=\vsize
\epsfbox{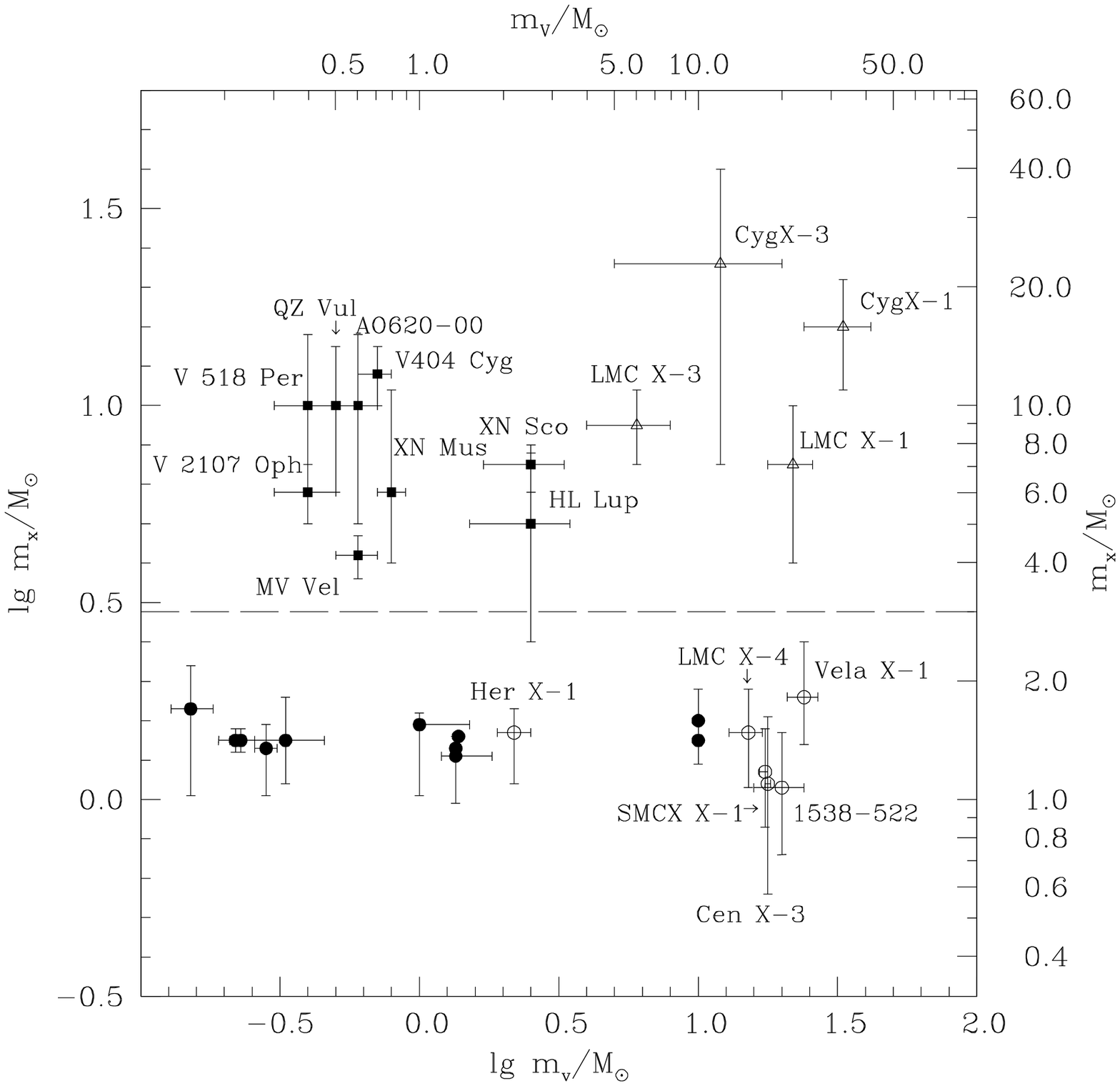}
}
\caption{Dependence of the masses $m_x$ of NS (circles)
and BH (triangles and rectangles) on the masses of their 
companion stars $m_V$ in close binary systems. Filled
circles correspond to radio pulsars, filled triangles to BH in 
X-ray novae.}
\label{rem_mas}
\end{figure}

\vskip\baselineskip
(2) {\bf Bimodality of GRB energy distribution and stellar remnant mass
distribution}.
The known masses of NS and BH in binaries are shown in Fig.
\ref{rem_mas}. NS masses are groupped ina anarrow interval is 
$M_{NS}=(1.35\pm 0.15) M_\odot$, while BH masses fall in a 
broader range
$M_{BH}=(5\div 15) M_\odot$. A real gap between NS and BH masses
is observed. 
 
GRBs with low energetics associated with peculiar supernovae
type Ic (such as GRB980425) can be explained by collapses of bare CO cores
of massive stars with significant rotation which causes most envelope to be
ejected and neutron star to be formed, while collapses of slower rotating
cores do not accompanied by a significant envelope ejection and lead to
black hole formation. In the latter case an energetic GRB can be generated
with energy proportional to the pre-collapse core mass.
\vskip\baselineskip  

(3) {\bf Association of GRBs 
with star-forming regions} 

\vskip\baselineskip
(4) {\bf A diversity of the observed afterglows}
A GRB in a binary system can induce different optical
phenomena due to illumination of the companion's atmosphere
by hard X-ray and gamma-radiation \cite{Blinnikov&Postnov97}. 
This should add some light
to the "pure" power-law afterglow from relativistic blast wave
thus producing a great variety of the observed light curves.  
These effects can occur in a time interval   
$\Delta t_{opt}>D/c$ after the burst 
($D$ is the distance to the optical star from GRB, $c$ is the 
speed of light). 
Deviations are indeed observed
in some bursts (for example, in GRB980326 afterglow 
three weeks after the burst \cite{JSBloom_ea99}).

Astronomical observations indicate \cite{Cherepashchuk01, Cherepashchuk98,
Cherepashchuk2000} that about 50\% of all WR stars in our Galaxy
can be in binaries with O-star or A-M-star.  
For example, for WR+O system V444 Cyg with an orbital period 
of $P=4^d.2$ we have $D\approx 40 R\odot$ and the time delay 
$\Delta t_{opt}\approx 100$~s, and for 
parameters of WR+O binary system 
CV Ser $\Delta t_{opt}\approx 300$~s.
An extremely bright optical emission ($V\simeq 9^m$) 
was observed in the famous burst GRB990123 
only 50 s after the burst beginning \cite{Akerlof_ea99}.

Another example is a peculiar shape of achromatic optical afterglow light
curve observed in GRB000301c
\cite{Masetti_ea00,Sagar_ea}. 
The observed several peaks  separated by 2-3 days 
days can be a manifestation of an 
orbital period in the underlied binary system, for example, through the
binary-period shaped mass loss before collapse. An alternative explanation
by a microlensing event \cite{Garnavich_ea00} seems less probable. 
Orbital periods of order of several days perfectly fit 
the observed period range
$1^d.6\div 2900^d$ in WR+O binary systems (see Table 2 in
\cite{Postnov&Cherepashchuk00}).

These arguments favor the GRB-WR stars association, but
there is a general requirement which should be met by all 
viable GRB models. The point is that GRB phenomenon should be an
extremely rare astronomical event. 
  
\section*{Event rate problem}

The GRB event rate per
unit comoving volume 
using BATSE data with $F_{tr}=0.1$ ph/cm$^2$ \cite{Stern_ea00} is 
$$
{\cal R}_{GRB}\sim 10^{-9} \hbox{GRB/yr/Mpc}^3\,,
$$
i.e about $10^{-7}$ per year in the average galaxy 
with a mass of $10^{11} M_\odot$. This is by several orders of magnitude
lower than the total rate of core collapses associated with 
SN II and Ibc
(${\cal R}_{SNIbc}\sim 3 \times 10^{-5}$/yr/Mpc$^3$,
\cite{Cappellaro_ea97}). This discrepancy is usually eliminated 
by introducing a beaming of gamma-ray emission   
(e.g. \cite{Lamb00}). It is not excluded that not each SN Ibc
is associated with GRB for internal reasons. 

The mean formation rate 
of all types of WR stars in the Galaxy is
\begin{equation}
{\cal R}_{WR} = {\cal} R_\odot\frac{N_{WR}}{N_\odot}
\frac{\Delta t_\odot}{\Delta t_{WR}}\sim \frac{1}{1000} \hbox{yr}^{-1}\,.
\end{equation}
i.e. by a factor of 1000 exceeds that of GRBs. The most compact 
WR stars, so called WO stars, are much less frequent 
(3 out of total 200 are known in the Galaxy), so their
formation rate is only by one order of magnitude higher than 
that of GRB.  
This issue can be solved either by
postulating generically thin jets or, admitting quasi-spherically symmetric
emission, by assuming the existence of some "hidden" collapse parameters
(rotation, magnetic field, etc.), which was suggested by  
\cite{Ergma&vdH98} from an
independent analysis of black hole formation in binaries. 

In the hypernova
scenario by Paczy\'nski \cite{Paczynski97} the rarity of GRB phenomenon is
also explained by requiring  an extremely high magnetic field during 
core collapse of a rotating massive star into a 10 $M_\odot$ black hole.

In contrast, in the model of coalescing neutron star/black hole binaries
(which is currently less favored by association of all observed GRB hosts
with strong star forming regions, see above) the event rates varies from
$\sim 10^{-4}$ to $\sim 10^{-6}$ per year depending on the binary 
evolution parameters \cite{LPP97}, which is marginally consistent
with the observed GRB rate and the event rate problem 
is not very strong. 

\section*{GRBs as a transient galactic phenomenon}

There is another possibility  to explain the observed association of cosmic
GRBs with star-forming regions at high redshifts and their extreme rarity.
GRBs may represent a transient galactic phenomenon occurring at the early
stages of galactic evolution, like quasars and AGNs. It is established now
\cite{Madau_ea96} that at high redshifts $z\sim 1-2$ a violent epoch of star
formation in young galaxies occurred. It is also known that a lot of cold
matter were bound in giant proto-galactic clouds at redshifts $z>2$, which
are observed as "Lyman-alpha forest" of absorption lines in quasar spectra.
The formation of very massive stars 100-500 $M_\odot$ which final 
collapse into massive
black holes took place at that epoch. Such 
massive star can not form from matter enriched with metals
because of pulsational instabilities (see \cite{Baraffe_ea00aph/009410}
and references therein). 
At low metallicity at the epoch of violent
star formation, however, they could have formed. The 
possibility of energetic GRBs from collapses of such massive stars
was studied in \cite{Fryer_ea00aph/0007176} with negative conclusion 
about their ability to produce an energetic GRB. But we note here that 
physical processes in such stars are still far from full 
understanding and potentially such stars could be 
GRB progenitors. The weakness of GRB980425 in a nearby
galaxy can be a natural consequence of smaller upper masses of  
stars in regions of violent star formation at 
the present epoch.

\end{document}